\documentclass[a4paper, 10pt, twocolumn]{article}

\usepackage[ansinew]{inputenc}          
\usepackage{graphicx}               
\usepackage{tabularx}               
\usepackage{multirow}               
\usepackage{url}                

\usepackage[small,bf]{caption2}     

\usepackage{parskip}

\usepackage{titlesec}
\usepackage{amsmath}                

\titleformat{\section}{\normalfont\large\bfseries}{\thesection}{}{}
\titleformat{\subsection}{\normalfont\large\bfseries}{\thesection}{}{}
\titleformat{\paragraph}{\normalfont\bfseries}{\theparagraph}{}{}
\titlespacing{\section}{0pt}{6pt}{-1pt}
\titlespacing{\subsection}{0pt}{3pt}{-1pt}
\titlespacing{\paragraph}{0pt}{3pt}{-1pt}

\newcolumntype{Y}{>{\centering\arraybackslash}X}    

\begin{document}

\date{}                                         

\title{\vspace{-8mm}\textbf{\large
Acoustics of Margravial Opera House Bayreuth }}  

\author{Sebastian Krauss, Simeon V\"olkel, Christoph Dobner, Alexandra V\"olkel, and 
Kai Huang$^1$\\
\emph{\small Experimentalphysik V, Universit\"at Bayreuth, 95440 Bayreuth, Germany} \\
$^1$\emph{\small Email: kai.huang@uni-bayreuth.de } }\maketitle

\begin{abstract}

The Margravial Opera House Bayreuth, built between 1745 and 1750, is a well preserved Baroque court theatre designed by Giuseppe Galli Bibiena~\cite{UNESCO}. It provides an opportunity to experience not only the visual but also the acoustic design of opera theatres in the 18th century, as the bell-shaped auditorium along with the decoratively painted canvas remains intact. Using balloons and hand-claps as sound sources, we characterize the impulse response of this opera house after its recent renovation. The reverberation time (RT), early decay time (EDT) and clarity factor are characterized and discussed in comparison to historical Italian theatres of a similar age.  

\end{abstract}

\thispagestyle{empty}           

\section*{Introduction}
\label{sec:intro} 

When the Margravial Opera House Bayreuth was being inscribed in the UNESCO's list of world cultural and natural heritage in 2012, one justification for its outstanding value was that ``The Margravial Opera House is a masterwork of Baroque court theatre architecture by Giuseppe Galli Bibiena in terms of its tiered loge form and acoustic, decorative and iconological properties.''~\cite{UNESCO} While the extraordinary interior decorations along with the 18th Baroque fa\c{c}ade are readily perceivable through site tours and photographs, the acoustics of this opera house is far less discussed. Motivated by this fact, following a recent characterization of the Bayreuth Festspielhaus~\cite{Huang2017}, we measure the impulse response of this opera house to provide an objective evaluation of its acoustics. This measurement is conducted roughly half a year after the recent renovation.

Below is a short history of the Opera House adapted from the book edited by Rainer~\cite{Rainer2018}, with an emphasis on the main changes made to the opera house in the past 270 years. This court opera house was originally built to celebrate the marriage of the only daughter of Margrave Friedrich and Margravine Wilhelmine in 1748. Since then, it had been regularly used by Wilhemine, who enjoyed writing librettos as well as composing music, despite of the high maintaining cost. After the death of Wilhemine in 1758, the Margravial opera house had rarely been used and hence kept largely intact until 1817, when the first renovation took place. The depth of the stage area for performance was reduced by roughly 45$\%$. Warm air heating systems, lightening and a fire protection system have been subsequently installed as the opera house became more and more often used as a municipal theatre. It was this opera house that attracted Richard Wagner to Bayreuth for implementing his festival idea in 1871, although he later considered that this opera house should better be kept exactly as it was. During the second renovation (1935-1936), the proscenium area had been modified, including the replacement of the wooden railing with a curving balustrade, new stairs in front of the stages, as well as a reduced size of stage opening. Those changes had been reversed to restore the original substance in the most recent renovation (2013-2018), during which roughly 90$\%$ of the original painting layer had been recovered, stabilized and strengthened. To summarize, beside the non-original wooden floor, the stage set and possible modifications to the proscenium area, the acoustics of the opera house nowadays should be very close to its original status in the 18th century.

\section*{Measuring Procedure}
\label{sec:method}

We characterize the acoustics of the Margravial opera house by measuring the impulse response, based on which further characterizations on the strength, music clarity, speech intelligibility, echoes, reverberation, and other features of a room can be subsequently quantified \cite{Kuttruff2000,Mueller2012}. Nowadays, such characterizations are becoming more convenient using hand claps and embedded MEMS (micro-electro-mechanical systems) microphones in smart phones \cite{Seetharaman2012,Rosenkranz2017}. In a previous investigation, it was demonstrated that hand-claps and smart phone recording (HCSP) approach can be used to characterize the reverberation time of a room reasonably well, except for the low frequency regime~\cite{Huang2017}. Here, in addition to HCSP, air balloons and condenser microphones are also used as sound sources and recording devices, respectively. The recorded signals are consequently analyzed with Matlab using the ITA-Toolbox \cite{Dietrich2013}, which provides a standard routine for room acoustic characterizations following ISO 3382 \cite{iso}.

\begin{figure}[hbt]
    \begin{center}
        \hskip 2em
		\hspace{-2em}\includegraphics[width=0.95\columnwidth]{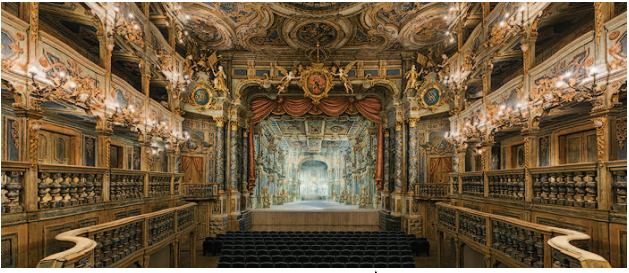} \\
		\includegraphics[width=0.95\columnwidth]{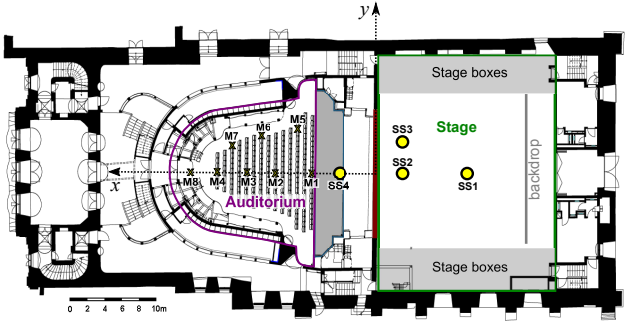}
    \end{center}
    \caption{Upper panel: The interior of the Margravial opera house viewed from the court loge~\cite{OperahouseSite}. Lower panel: Ground plan of the Margravial Opera House Bayreuth adapted from Ref.~\cite{Rainer2018} with the borders of the auditorium, stage, as well as the locations of the sound sources (SS) and recording microphones (M) marked. M8 is located in the court loge opposite to the stage. The gray shaded region marks orchestra area. See Table~\ref{tab:pos} for measured coordinates in the Cartesian system defined by dashed arrow lines. The thick red line corresponds to a iron curtain, which was closed for measurement SS4 and open for SS1-3.}
    \label{fig:method}  
\end{figure}

As shown in Fig.~\ref{fig:method}, the Margravial opera house in Bayreuth has a bell shaped auditorium including stalls, a gallery and three tiers of loges. This design resembles the plan of Lamberti's ``ideal'' theatre~\cite{Barbieri1998}. The heavily decorative ornaments are designed on the one hand for the festival, and on the other hand for improving the acoustics inside, representing the belief of the architects~\cite{Barbieri1998}. Except for the floor, the auditorium retains its original building material, i.e., wood~\cite{Rainer2018}. More specifically, linden (tilia) was used for the sculptures, while fir (abies) and spruce (picea) were used for the other parts. Part of the wooden walls and the ceiling are covered with canvas. The auditorium has a total volume of about $5000$m$^3$ with 520 seats distributed on the floor, three tiers of loges, the court loge and the gallery. 


\begin{table}[htbp]
    \centering
    \caption{Positions (in meter) of sound sources and recording devices in the coordination system defined in Fig.\,\ref{fig:method}(b).}
    \vspace{2mm}
    \label{tab:pos}
        \begin{tabularx}{8.5cm}{@{\arrayrulewidth1.5pt\vline}Y@{\arrayrulewidth1.5pt\vline} Y|Y|Y|Y|Y|Y|Y|Y @{\arrayrulewidth1.5pt\vline}}
            \noalign{\hrule height1.5pt}  & \multicolumn{2}{c@{\arrayrulewidth1.5pt\vline}}{SS1}
			& \multicolumn{2}{c@{\arrayrulewidth1.5pt\vline}}{SS2}
			& \multicolumn{2}{c@{\arrayrulewidth1.5pt\vline}}{SS3}
			& \multicolumn{2}{c@{\arrayrulewidth1.5pt\vline}}{SS4}\\\hline
			X  & \multicolumn{2}{c@{\arrayrulewidth1.5pt\vline}}{-17.7}
			& \multicolumn{2}{c@{\arrayrulewidth1.5pt\vline}}{-2.7}
			& \multicolumn{2}{c@{\arrayrulewidth1.5pt\vline}}{-2.7}
			& \multicolumn{2}{c@{\arrayrulewidth1.5pt\vline}}{4.1}\\\hline
			Y  & \multicolumn{2}{c@{\arrayrulewidth1.5pt\vline}}{0}
			& \multicolumn{2}{c@{\arrayrulewidth1.5pt\vline}}{0}
			& \multicolumn{2}{c@{\arrayrulewidth1.5pt\vline}}{2.4}
			& \multicolumn{2}{c@{\arrayrulewidth1.5pt\vline}}{0}\\
            \noalign{\hrule height1.5pt}  & M1 & M2 & M3 & M4 & M5 & M6 & M7 & \hspace{-0.7em} M8 \\
			\hline X & 6.5 & 10.7 & 14.9 & 19.1 & 7.8 & 12.3 & 17.5 & \hspace{-0.5em}23.5 \\
			\hline Y & 0 & 0 & 0 & 0 & 4.1 & 3.3 & 2.4 &  0  \\
            \noalign{\hrule height1.5pt}
        \end{tabularx}
\end{table}

The measurement was conducted on October 11, 2018. In the auditorium, the temperature and relative humidity in the auditorium was 18.9 degrees and 45$\%$, respectively. Impulse signals are generated either by air balloons (AB) or hand-claps (HC). The balloons are inflated with a pump to a diameter of 23$\pm 3$\,cm (measured at the equator) and pricked with a needle (outer and inner diameter $2$\,mm and $1.4$\,mm, respectively) to generate impulse signals. Both AB and HC signals are released from a height of $\sim$1.8\,m above the ground. Two room conditions with four sound source (SS) locations have been chosen: Iron curtain closed (SS4) or open (SS1-3). At each SS location, 10 AB followed by 10 HC signals are generated and subsequently recorded by the microphones distributed in the auditorium. Six smart-phones are clamped to the chairs with flexible arms and oriented to face the stage. The levels of the microphone gain are adjusted before each measurement and the automatic gain control is disabled. In addition, three condenser microphones (one Superlux ECM999 and two AKG C1000S) are also used as recording devices. Except for the combination of SS4 with M1, the distance between each sound source and recorder is at least twice the reverberation distance~\cite{Mueller2012}. The sampling rate for all recordings is $44.1$~kHz. The recorded signals from all microphones are automatically cut into individual impulses. Using the ITA-tool box, the raw signals are filtered into different octave bands and subsequently the energy decay curves for each band (EDC) are obtained through applying the backward integration method on the squared envelope of the filtered signals~\cite{Kuttruff2000}. From the least square fits of individual EDCs in a semi-logarithmic plane, the reverberation times are obtained. Depending on the different ranges of data used in the fitting, EDT (early reverberation time), T15, T20, T30, etc., are obtained. The uncertainty for the RT is estimated from the residuals around the regression line~\cite{Higbie1978}, taking into account their autocorrelation~\cite{Campbell1977,Moulton1986} arising from the backward integration method. The autocorrelation is approximated by its upper bound $1$ to have a conservative error estimation. In the present study, EDT, RT and clarity index C80 are obtained in the frequency range from $125$~Hz to $4000$~Hz, covering six octave bands. 


\section*{Results and Discussion}
\label{sec:ir} 

\begin{figure}[h]
    \begin{center}
		\includegraphics[width=1\columnwidth]{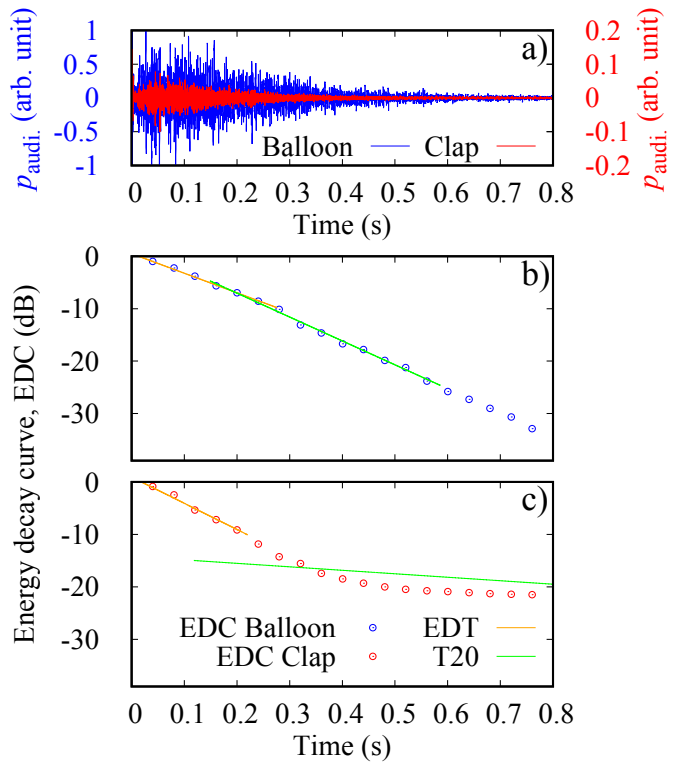}
    \end{center}
    \caption{ (a) Representative raw signals from one balloon-burst and one hand-clap at SS3 recorded by M3. (b, c) Corresponding energy decay curves of the impulse from the bursting balloon and from the hand-clap at the 1000\,Hz octave band. The orange and green solid lines correspond to linear fits for obtaining the early decay time and T20, respectively. For the EDC curves, only every 600th data point is shown for a better visibility.}
    \label{fig:raw}  
\end{figure}

Figure~\ref{fig:raw} shows a comparison of two sample raw signals generated by a balloon-burst as well as by a hand-clap. It shows that the sound pressure level (SPL) generated by hand-claps is roughly one order of magnitude smaller than that from balloon-bursts. Consequently, the EDC curve in the former case decays rather quickly to the surrounding noise level in comparison to the latter case. Therefore, the characterization of RT with T20 or above becomes unfeasible, as the fitted curve for T20 in (c) shows. In order to obtain reliable RTs, we set a threshold for the uncertainty obtained from a least square fit: Only RTs with a relative uncertainty smaller than $10\%$ are used in the following analysis. With this criterion, unrealistic RTs [e.g. fit in (c) delivers T20=$5.2\pm1.9$\,s] can be sorted out. Only data obtained from balloon experiments are used in the following characterization, because the impulses generated from bursting balloons are more reliable concerning repeatability~\cite{Paetynen2011} and feasibility.  

\begin{figure}[hbt]
    \begin{center}
	\includegraphics[width=0.95\columnwidth]{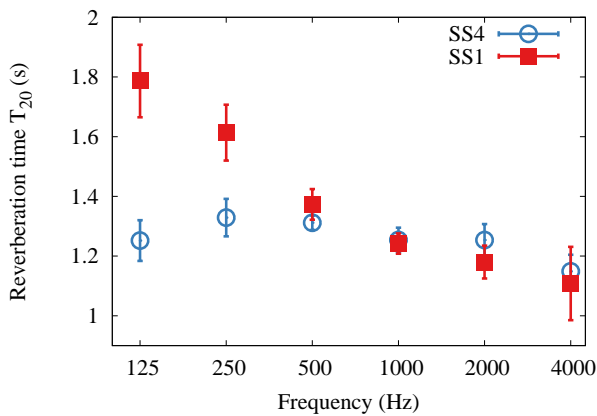}
    \end{center}
    \caption{Reverberation time (T20) of the auditorium at different octave bands for two different conditions: With the iron curtain open (SS1) or closed (SS4). It is averaged over all microphones except for M2, which has technical issues. The error bars correspond to the maximum of the uncertainty from fitting and that from data scattering among different microphones.}
    \label{fig:rt}  
\end{figure}

The spectral variation of the reverberation time is shown in Fig.\,\ref{fig:rt} and listed in Tab.\,\ref{table:rt}. Similar to `regular' Italian historical opera houses, there exists a clear increase of RT as frequency decreases to the 125 and 250~Hz octave bands, arising from the extensive mid- and high-frequency sound absorptions due to the interior design~\cite{Prodi2015}. For instance, the loges act as resonators for low frequency sound, and meanwhile as effective absorbers for mid- and high-frequency sound, owing to the decorative design and sound absorbing finishes. For a better visibility, only T20 values are shown here as representatives. The other RT values agree with T20 within experimental uncertainties, provided that the fitting criterion described above is applied. The reverberation times of the auditorium for low (125 and 250\,Hz octave bands) and middle (500 and 1000\,Hz octave bands) frequency ranges, averaged over various sound sources and microphones are $1.67\pm0.11$\,s and $1.29\pm0.02$\,s, respectively. This results lead to a bass ratio (BR) of $1.29$, matching the grand average (1.30 in unoccupied conditions) of 50 traditional Italian opera houses characterized recently by Prodi \textit{et al.}~\cite{Prodi2015}. 

When the iron curtain is closed (condition for measurements taken at SS4), the overall reverberation time reduces to $1.29\pm0.05$\,s and $1.28\pm0.03$\,s for low- and middle-frequencies, respectively. According to the Sabine formula~\cite{Sabine1922, Kuttruff2000}, RT $\propto V/S$ with total volume $V$ and surface area $S$ of the room. As $V$ is dramatically reduced with closed iron curtain, and in the meanwhile $S$ does not change with the same proportionality as $V$ (due to the decorative, sound absorbing surfaces), the overall RT is expected to decrease substantially. The other obvious feature is the vanishing spectral variation, which arises presumably from the vanishing coupling between the auditorium and the stage~\cite{Kuttruff2000}.


\begin{table}[htbp]
    \centering
    \caption{Averaged reverberation time obtained without (SS1) and with
(SS4) ironcurtain closed, unoccupied.}
    \vspace{2mm}
    \label{table:rt}
        \begin{tabularx}{8.5cm}{@{\arrayrulewidth1.5pt\vline}Y@{\arrayrulewidth1.5pt\vline}Y|Y|Y|Y|Y|Y|Y
@{\arrayrulewidth1.5pt\vline}}
            \noalign{\hrule height1.5pt}  & 125 & 250 & 500 & 1000 & 2000
& 4000  &Hz \\
                        \hline SS1 & 1.79 & 1.61 & 1.37 & 1.24	& 1.18 &
1.11 & s \\
                        \hline SS4 & 1.25 & 1.33 & 1.31 & 1.25 & 1.25 &
1.15 & s \\
            \noalign{\hrule height1.5pt}
        \end{tabularx}
\end{table}

To estimate the transparency of music in this opera house, we also characterize the clarity index C80~\cite{Kuttruff2000}, which characterizes the relative importance of the early part (within $80$\,ms) with respect to the later part of the impulse response, and hence provides information on the effectiveness of sound reflections. For all the combinations of SS and M, C80 ranges from 0.5 to 3.2. The average value of C80 in the stalls is $1.9\pm0.6$ and falls in the preferred range by Hidaka and Beranek~\cite{Hidaka2000}. It suggests that the subjective clarity is sufficient even for fast musical passages. Together with the averaged EDT $\sim 1.12\pm0.15$\,s (with iron curtain open), we can compare the Margravial opera house with traditional Italian theatres as well as modern theatres after the work of Prodi \textit{et al.}~\cite{Prodi2015}. It is interesting to see that the Margravial opera house falls into the group of ``regular'' theatres without large hard reflecting surfaces. Although C80 has a clear spatial distribution in the auditorium, the relatively small standard deviation $0.6$ suggests that the overall good transparency and clarity for music, at least in the stalls that were characterized.

\section*{Conclusions and Outlook}
\label{sec:sum} 

To summarize, room acoustic characterizations are conducted for the newly renovated Margravial opera house in Bayreuth. Our investigation suggests that the acoustics of this opera house matches perfectly that of typical ``Italian-style historical opera houses'' built between 1637 and 1887~\cite{Prodi2015}, representing the belief of the architects on ``ideal'' theatres~\cite{Barbieri1998}. It has a relatively low reverberation time in comparison to modern theatres due to the decorative interior design and the lack of large hard reflecting surfaces. It delivers fairly well clarity at least for the stalls. For the loges and boxes, C80 is expected to be influence by detailed configurations of seating in the boxes. Therefore, further characterizations in the loges, particularly with the presence of audience are needed to understand its acoustic design better.

From a technical perspective, this investigation shows that bursting balloons serve as a more repeatable handy sound source in comparison to hand-claps. For recorded signals with relatively weak signal-to-noise ratio, setting up a threshold for the goodness of fit is valuable for obtaining more reliable and reproducible reverberation times.

\section*{Acknowledgments}

We thank Thomas Rainer from the `Bayerische Schl\"osserverwaltung' and Angela Danner from the press office of Bayreuth University for bringing us the opportunity of conducting the measurement. The detailed information on the history and technical aspects of the opera house kindly provided by Thomas Rainer helped greatly in conceiving the original acoustic design of the theatre. SK acknowledges support by the Elite Network of Bavaria (Study Program Biological Physics). We are also grateful to Tobias Eckert, Christoph Schnupfhagen, Nico Stuhlm\"uller, and Michael Seidel for their kind help in audio recording.                                        


\begin{thebibliography}{10}

\bibitem{UNESCO}
\newblock UNESCO World~Heritage Centre \\
\newblock https://whc.unesco.org/en/list/1379/ [accessed 2019-02-01].

\bibitem{Huang2017}
K.~Huang.
\newblock Impulse response of the bayreuth festspielhaus.
\newblock In {\em Fortschritte der Akustik}, pages 238--241, Kiel, Germany,
  2017.

\bibitem{Rainer2018}
Thomas Rainer.
\newblock {\em Margravial Opera House Bayreuth}.
\newblock Bayerische Schloesserverwaltung, 2018.

\bibitem{Kuttruff2000}
Heinrich Kuttruff.
\newblock {\em Room Acoustics}.
\newblock CRC Press, London, England ; New York, NY, 4th edition edition,
  October 2000.

\bibitem{Mueller2012}
Gerhard M\"uller and Michael M\"oser.
\newblock {\em Handbook of Engineering Acoustics}.
\newblock Springer, Berlin, Heidelberg, November 2012.

\bibitem{Seetharaman2012}
Prem Seetharaman and Stephen~P. Tarzia.
\newblock The hand clap as an impulse source for measuring room acoustics.
\newblock In {\em Audio Engineering Society Convention 132}, Apr 2012.

\bibitem{Rosenkranz2017}
Andreas Rosenkranz, Ralf Burgmayer, David Ackermann, Markus H\"adrich, and
  Stefan Weinzierl.
\newblock Eine mobile app zur bestimmung raumakustischer parameter.
\newblock In {\em the 43th German Annual Conference on Acoustics (DAGA)}, 2017.

\bibitem{Dietrich2013}
Pascal Dietrich, Martin Guski, Johannes Klein, Markus M\"uller-Trapet, Martin
  Pollow, Roman Scharrer, and Michael Vorl\"ander.
\newblock Measurements and room acoustic analysis with the ita-toolbox for
  matlab.
\newblock In {\em 40th Italian (AIA) Annual Conference on Acoustics and the
  39th German Annual Conference on Acoustics (DAGA)}, 2013.

\bibitem{iso}
ISO3382-1, acoustics measurement of room acoustic parameters. part 1:
  Performance spaces, 2009.

\bibitem{OperahouseSite}
\newblock http://www.bayreuth-wilhelmine.de [accessed 2019-02-01]. 

\bibitem{Barbieri1998}
Patrizio Barbieri.
\newblock The acoustics of italian opera houses and auditoriums (ca.
  1450-1900).
\newblock {\em Recercare}, 10:263--328, 1998.

\bibitem{Higbie1978}
J.~Higbie.
\newblock Uncertainty in a least-squares fit.
\newblock {\em American Journal of Physics}, 46(9):945--945, September 1978.

\bibitem{Campbell1977}
Cathy Campbell.
\newblock Properties of ordinary and weighted least square estimators of
  regression coefficients for two-stage samples.
\newblock In {\em Proceedings of the {Social} {Statistics} {Section},
  {American} {Statistical} {Association}}, pages 800--805, 1977.

\bibitem{Moulton1986}
Brent~R. Moulton.
\newblock Random group effects and the precision of regression estimates.
\newblock {\em Journal of Econometrics}, 32(3):385--397, August 1986.

\bibitem{Paetynen2011}
Jukka P\"atynen, Brian~F.G. Katz, and Tapio Lokki.
\newblock Investigations on the balloon as an impulse source.
\newblock {\em The Journal of the Acoustical Society of America},
  129(1):EL27--EL33, January 2011.

\bibitem{Prodi2015}
Nicola Prodi, Roberto Pompoli, Francesco Martellotta, and Shin-ichi Sato.
\newblock Acoustics of {Italian} {Historical} {Opera} {Houses}.
\newblock {\em The Journal of the Acoustical Society of America},
  138(2):769--781, August 2015.

\bibitem{Sabine1922}
W.~C. Sabine.
\newblock {\em Collected papers on acoustics}.
\newblock Harvard University Press, 1922.

\bibitem{Hidaka2000}
Takayuki Hidaka and Leo~L. Beranek.
\newblock Objective and subjective evaluations of twenty-three opera houses in
  {Europe}, {Japan}, and the {Americas}.
\newblock {\em The Journal of the Acoustical Society of America},
  107(1):368--383, December 2000.

\end{thebibliography}

\end{document}